\documentstyle[12pt]{article}

\def\d{\partial}
\def\2{\frac{1}{2}}
\def\z{\bar z}
\def\w{\bar w}
\def\x{\bar x}
\def\a{\alpha}
\def\b{\beta}
\def\g{\gamma}

\def\bb{\bar \beta}
\def\bg{\bar \gamma}
\def\p{\phi}
\def\Scr{\mathop{\rm Scr}\nolimits}

\def\dim{\mathop{\rm dim}\nolimits}
\def\const{\mathop{\rm const}\nolimits}
\def\beq{\begin{equation}}
\def\eeq{\end{equation}}
\def\ba{\beq\new\begin{array}{c}}
\def\bal{\begin{array}{l}}
\def\eal{\end{array}}
\def\bac{\begin{array}{c}}
\def\eac{\end{array}}

\def\ea{\end{array}\eeq\\}
\def\be{\ba}
\def\ee{\ea}

\makeatletter
\newdimen\normalarrayskip              
\newdimen\minarrayskip                 
\normalarrayskip\baselineskip
\minarrayskip\jot
\newif\ifold             \oldtrue            \def\new{\oldfalse}
\def\arraymode{\ifold\relax\else\displaystyle\fi} 
\def\eqnumphantom{\phantom{(\theequation)}}     
\def\@arrayskip{\ifold\baselineskip\z@\lineskip\z@
     \else
     \baselineskip\minarrayskip\lineskip2\minarrayskip\fi}
\def\@arrayclassz{\ifcase \@lastchclass \@acolampacol \or
\@ampacol \or \or \or \@addamp \or
   \@acolampacol \or \@firstampfalse \@acol \fi
\edef\@preamble{\@preamble
  \ifcase \@chnum
     \hfil$\relax\arraymode\@sharp$\hfil
     \or $\relax\arraymode\@sharp$\hfil
     \or \hfil$\relax\arraymode\@sharp$\fi}}
\def\@array[#1]#2{\setbox\@arstrutbox=\hbox{\vrule
     height\arraystretch \ht\strutbox
     depth\arraystretch \dp\strutbox
     width\z@}\@mkpream{#2}\edef\@preamble{\halign
\noexpand\@halignto
\bgroup \tabskip\z@ \@arstrut \@preamble \tabskip\z@ \cr}%
\let\@startpbox\@@startpbox \let\@endpbox\@@endpbox
  \if #1t\vtop \else \if#1b\vbox \else \vcenter \fi\fi
  \bgroup \let\par\relax
  \let\@sharp##\let\protect\relax
  \@arrayskip\@preamble}
\def\eqnarray{\stepcounter{equation}%
              \let\@currentlabel=\theequation
              \global\@eqnswtrue
              \global\@eqcnt\z@
              \tabskip\@centering
              \let\\=\@eqncr
              $$%
 \halign to \displaywidth\bgroup
    \eqnumphantom\@eqnsel\hskip\@centering
    $\displaystyle \tabskip\z@ {##}$%
    \global\@eqcnt\@ne \hskip 2\arraycolsep
         $\displaystyle\arraymode{##}$\hfil
    \global\@eqcnt\tw@ \hskip 2\arraycolsep
         $\displaystyle\tabskip\z@{##}$\hfil
         \tabskip\@centering
    &{##}\tabskip\z@\cr}
\begingroup\ifx\undefined\newsymbol \else\def\input#1 {\endgroup}\fi
\newfont{\hr}{msbm10}
\newfont{\ams}{msam10}

\font\numbers=cmss12
\font\upright=cmu10 scaled\magstep1
\def\stroke{\vrule height8pt width0.4pt depth-0.1pt}
\def\topfleck{\vrule height8pt width0.5pt depth-5.9pt}
\def\botfleck{\vrule height2pt width0.5pt depth0.1pt}
\def\Zmath{\vcenter{\hbox{\numbers\rlap{\rlap{Z}\kern 0.8pt\topfleck}\kern
2.2pt
                   \rlap Z\kern 6pt\botfleck\kern 1pt}}}
\def\Qmath{\vcenter{\hbox{\upright\rlap{\rlap{Q}\kern
                   3.8pt\stroke}\phantom{Q}}}}
\def\Nmath{\vcenter{\hbox{\upright\rlap{I}\kern 1.7pt N}}}
\def\Cmath{\vcenter{\hbox{\upright\rlap{\rlap{C}\kern
                   3.8pt\stroke}\phantom{C}}}}
\def\Rmath{\vcenter{\hbox{\upright\rlap{I}\kern 1.7pt R}}}
\def\Z{\ifmmode\Zmath\else$\Zmath$\fi}
\def\Q{\ifmmode\Qmath\else$\Qmath$\fi}
\def\N{\ifmmode\Nmath\else$\Nmath$\fi}
\def\C{\ifmmode\Cmath\else$\Cmath$\fi}
\def\R{\ifmmode\Rmath\else$\Rmath$\fi}
\def\numberbysection{\@addtoreset{equation}{section}
        \def\theequation{\thesection.\arabic{equation}}}
\numberbysection

\renewcommand{\theequation}{\thesection.\arabic{equation}}
\newcommand{\l@qq}[2]{\addvspace{2em}
 \hbox to\textwidth{\hspace{1em}\bf #1 \dotfill #2}}

\newcounter{app}

\def\app{\setcounter{equation}{0}
\def\theequation{\Alph{app}.\arabic{equation}}\par
   \addvspace{4ex}
   \@afterindentfalse
  \secdef\@app\@dapp}
\newcommand\@app{\@startsection {app}{1}{0ex}%
                                   {-3.5ex \@plus -1ex \@minus -.2ex}%
                                   {2.3ex \@plus.2ex}%
                                   {\normalfont\Large\bf}}

\def\@dapp#1{%
{\parindent \z@ \raggedright  \bf #1}\par\nobreak}
\def\l@app#1#2{\ifnum \c@tocdepth >\z@
    \addpenalty\@secpenalty
    \addvspace{1.0em \@plus\p@}%
    \setlength\@tempdima{2.5em}%
    \begingroup
      \parindent \z@ \rightskip \@pnumwidth
      \parfillskip -\@pnumwidth
      \leavevmode \bfseries
      \advance\leftskip\@tempdima
      \hskip -\leftskip
      #1\nobreak\hfil \nobreak\hb@xt@\@pnumwidth{\hss #2}\par
    \endgroup\fi}
\newcounter{sapp}[app]

\def\sapp{\def\theequation{\Alph{app}.\arabic{equation}}\par
   \@afterindentfalse
  \secdef\@sapp\@dsapp}
\newcommand\@sapp{\@startsection{sapp}{2}{\z@}%
                                     {-3.25ex\@plus -1ex \@minus -.2ex}%
                                     {1.5ex \@plus .2ex}%
                                     {\normalfont\large\bfseries}}

\def\@dsapp#1{%
{\parindent \z@ \raggedright  \bf #1}\par\nobreak}
\newcommand{\l@sapp}{\@dottedtocline{2}{1.5em}{3em}}

\def\der#1#2{\frac{\d{#1}}{\d{#2}}}

\def\rank{{\rm rank}}
\def\N2{${\cal N}=2$}

\def\theequation{\thesection.\arabic{equation}}

\begin{document}

\thispagestyle{empty}

\begin{flushright}
ITEP-TH-52/99\\
hep-th/9912042
\end{flushright}

\vspace{2.0cm}

\centerline{\LARGE  Conformal Blocks and Correlators in WZNW Model.}

\bigskip

\centerline{\LARGE I. Genus Zero. }

\vspace{0.5cm}

\bigskip

\setcounter{footnote}{0}

\setcounter{equation}{0}

\bigskip

\begin{center}

Kirill Saraikin\footnote{e-mail: saraikin@itp.ac.ru}\\

\bigskip

\bigskip

{\em L.D.Landau Institute for Theoretical Physics, 117334, Moscow,
Russia\\

and\\

Institute of Theoretical and Experimental Physics, 117259, Moscow,
Russia}

\end{center}

\bigskip

\bigskip

\begin{quotation}

We consider the free field approach or bosonization technique for the
Wess-Zumino-Novikov-Witten model with arbitrary Ka\v{c}-Moody algebra
on  Riemann surface of genus zero. This subject was much studied
previously, and the paper can be partially taken as a brief survey.
The way to obtain well-known Schechtman-Varchenko solutions of the
Knizhnik-Zamolodchikov equations as certain correlators in free chiral
theory is revisited. This gives rise to simple description of space of
the WZNW conformal  blocks. The general $N$-point correlators of the
model are constructed from the conformal blocks using non-chiral action
for free fields perturbed by exactly marginal terms. The method
involved generalizes the Dotsenko-Fateev prescription for minimal
models. As a consequence of this construction we obtain
new integral identities.

\end{quotation}

\newpage

\section{Introduction}

The Wess-Zumino-Novikov-Witten model~\cite{Nov,Wit, PW} has a long
history. Its exceptional role is determined by the fact
that this model can be in broad sense thought of as a generator of
all the 2d conformal theories~\cite{MS}. Since the seminal work by
Knizhnik and Zamolodchikov~\cite{KZ} much effort has been made to
obtain exact solution of WZNW
model~\cite{ZF,CFlu,BF,GMMOS,Dots,Awa,FGPP}.
The most progress was achieved in solving the model by means
of its representation in terms of the free fields (bosonization
approach). Since the
work by Wakimoto~\cite{Wak} this approach has been developed by many
groups~\cite{GMMOS,AS,FFr,BMcCP,PRY,PRY1}.

In fact, it is possible to think of conformal theories
as just of theories of free fields with additional
constraints on the space of states.
A good example of how this idea works is the well-known
Dotsenko-Fateev free field representation for minimal
models~\cite{DF}. There exist a lot of serious arguments in
favour of applicability of such a viewpoint to the WZNW model.
For instance, in geometrical quantization approach the WZNW Lagrangian
may be naturally considered as a $d^{-1}$ of a Kirillov-Konstant
form on the orbit of Ka\v{c}-Moody coajoint representation~\cite{AS}.
After choosing Gauss parameterization for the group element the
action becomes diagonalized and quadratic in the corresponding
fields~\cite{GMMOS}.
Thus the free field representation of the WZNW model
canonically arises.
Unfortunately, since in general the transformation to the
free variables is highly non-local, it is hard to describe
it carefully. In fact, this way one can only
conjecture expressions for the conformal blocks
(see~\cite{GMMOS} for more detailed discussion).
It is worth mentioning that there exists an alternative approach due
to Gaw\c{e}dzki et al.~\cite{Gaw} based on the correspondence
between the Chern-Simons theory in the bulk that spans some
surface and the WZNW model on this surface.
This correspondence links  the correlators in WZNW model and
scalar product on the quantum states space of the CS theory.
The last one can be calculated using the Iwasawa parameterization
of the gauge field.  After that the corresponding  action also
becomes Gaussian and explicit expressions for the scalar
products can be obtained~\cite{Gaw}.

However it seems that at present a brief and clear description
of such a point of view on the WZNW model
is absent in literature.
This paper is a small step in this direction.
Although the problem of primary interest is that of the WZNW model
on higher genus, we start our investigation  with the
simplest genus zero case.
This subject has been studied previously, and the paper can be
partially taken as a survey. Our other purpose here is to prepare
the necessary background for the second paper in the series,
devoted to the higher genus case.

Our ideology of the WZNW model description is quite simple.
The correlators and
hence conformal blocks should satisfy a set of the
operator product expansions (OPE),
fixed in bootstrap approach~\cite{KZ,BPZ}.
We realize these OPE in terms of free fields thus
ensuring in correct local properties of our expressions.
The correctness of the global properties follows then
from the useful fact that ``there are not so many good objects
on Riemann surface" (a slang variant of the Riemann-Roch theorem).
This rough statement turns to be remarkably confirmed:
as the result, we obtain the well-known
Schechtman-Varchenko solutions~\cite{SV} of the
Knizhnik-Za\-mo\-lod\-chi\-kov equations.

To represent all the primary fields from the multiplet we introduce
a generating function in a way which is quite similar to the
exponential map from algebraic to the group elements.
We also mention that one can use an alternative
generating function from~\cite{PRY,PRY1}, which corresponds to the
algebra representation on the polynomial ring.
However, to compare the answers with the solutions~\cite{SV}
one needs an exponential generating function.

Let us list some results:

{\bf 1.} For the $SU(2)_k$ WZNW model we claim that the $N$-point
correlator of the spinless primary fields  coincides with the
$N$-point correlator of the ``dressed" vertex operators
in the theory with the action

\be
\label{S'}
S_{\p \b \g}'=\frac{1}{4\pi} \int d^2z
\left\{
\2 \d \p \bar \d \p -\b \bar \d \g -\bb \d \bg +
i\sqrt{\frac{2}{k+2}} {\mathcal R} \p +
 \b \bb \exp \left(-i \sqrt{\frac{2}{k+2}} \p\right)
\right\}
\ee
The last term in the integrand is exactly marginal and one
should calculate correlators as a power series over this term.
In fact, often after functional integration a single
term survives in expansion.

The ``dressed" vertex operator associated with the
primary field from the highest weight $j$ representation is

\be
\tilde V_j = e^{\g f_L}  \
:e^{i \frac{\sqrt{2}}{q} j \p}:\
e^{\bg f_R}
\ee
where $f_L$ and $f_R$ are the $su(2)$ step generators, which form the
``left"  and ``right" representations.  Thus, our claim is:

\be
\left< \Phi_{\Delta _{j_1}} (z_1,\z_1)\dots
\Phi_{\Delta_{j_N}} (z_N,\z_N) \right>_{WZNW}=
 \const \ \left<\tilde V_{j_1}(z_1,\z_1) \dots \tilde V_{j_N}(z_N,\z_N)
\right>_{S_{\p \b \g}'}
\ee
This construction may be easily generalized for other
Lie groups.  As a result one can calculate the
$N$-point correlators in group $\bf G$ WZNW
model as correlators of ``dressed" vertex operators in the theory
of free fields perturbed by exactly marginal terms.
The number of the fields necessary for bosonization is
defined by the $\rank$  $\bf G$ while the marginal
terms correspond to  the ``squared modules" of simple
screening currents.

{\bf 2.} As a result of the above suggestion we obtain
expressions which
satisfy all the requirements necessary for the WZNW correlators:

i) the holomorphic factorization property

ii) correct conformal properties; this means
that the behavior of a correlator as a function of the
primary fields insertion points under the change of
coordinates on the surface is governed by the
stress-tensor and corresponding conformal dimensions

iii) satisfy the differential (Knizhnik-Zamolodchikov)
and additional algebraic equations, which reflects the
null-vectors decomposition

iv) be  a well-defined function of the primary fields
insertion points; in other words, the monodromy
of the correlator when one insertion
point is moved around the others should be trivial.

{\bf 3.} We suggest a set of  non-trivial integral
identities, the simplest one is the following:

\be
\label{ex}
|x|^{1/3} \ |1-x|^{1/3} \int d^2 t_1 \
|t_1|^{-2/3} \ |t_1-x|^{-2/3}\
|t_1-1|^{-2/3} \ \times\\
\times \int d^2 t_2 \ |t_1-t_2|^{4/3}\ |t_2|^{-2/3}
|t_2-x|^{-2/3}\ |t_2-1|^{-2/3}\ \left| \ \frac{1}{t_1(t_2-1)}+
\frac{1}{t_2(t_1-1)} \ \right|^2 = \\
= \frac{\const}{\sqrt{x \bar x (1-x)(1-\bar x)}}
\ee
All such identities occur in some exceptional cases, when an
alternative
way to solve the WZNW model exists. For instance, (\ref{ex})
corresponds to the $SU(2)_{k=1}$ case, when due to special
circumstances all correlators can be represented in terms of single
scalar bosonic field with values on self-dual circle, see section 5.
We have checked some of them numerically.
These tests ensure us in correctness of the
proposed construction.

The outline of this paper is as follows. In section~2 we fix notations
and briefly review the basics of the bosonization technique.

In section~3 we describe how the Schechtman-Varchenko solutions
of the Knizhnik-Za\-mo\-lod\-chi\-kov
equations can be obtained  as  certain correlators in the free chiral
theory. As a consequence, we obtain essential interpretation of the
``resonance conditions" from~\cite{FSV} as a neutrality condition
on vertex operators in the free field theory, which is necessary
for obtaining a non-vanishing result.

After that, in section~4 we turn to the problem of ``gluing"
correlators from the conformal blocks.
We use a simple prescription to obtain the conformal blocks:
First, add to the free (non-chiral) action exactly marginal terms
which are ``squared modules" of screening currents integrated over
whole surface in question. Second, take the power series over
these terms. The method involved generalizes the
Dotsenko-Fateev prescription for minimal models.

In section~5 we suggest a set of new integral identities.
These identities
arise when  an alternative way to solve the WZNW model exists.
The lhs and rhs are the correlators calculated in two different
ways.
They have equal conformal, algebraic and analytical properties.
Moreover,
they obey the same differential and algebraic equations. Thus at
least at
physical level we can conclude that they should be equal.
This results in the non-trivial relations between (generalized)
hypergeometric functions.
We have checked some of them numerically performing a number of
tests on
proposed construction.

Finally, section~6 describes various open problems and directions
for future research.

\section{Basics of the Bosonization Technique}

For most of the material presented in this section,
see~\cite{GMMOS,PRY}
and references therein. We concentrate our attention on the $su(2)$
algebra. After careful analysis of this case the generalization for
the arbitrary algebras is straightforward.

\subsection{Notations for the finite dimensional algebra }

Let $\bf g$ be a simple finite dimensional complex Lie algebra of
$\dim {\bf g}=d$ and $\rank {\bf g}=r$. Let $\left< \ ,\ \right> $
be an
invariant scalar product (Killing form) on  $\bf g$ normalized in
such a way
that $\left< \theta,\theta \right> =2$, $\theta$ being the highest
root. The
set of positive roots is denoted $\Delta_{+}$.  The simple roots are
$\{ \alpha_i \}_{i=1,\dots ,r}$.  For a vector space $\mathcal V$,\
$\mathcal V^{\vee}$ will always denote the dual space. The Cartan
matrix is
$A_{ij}=\left< \alpha_{i}^{\vee},\alpha_{j} \right> $, the dual
Coxeter
number is $h^{\vee}=\sum \limits_{i=1}^{r}\alpha_i^{\vee}+1$.
Commutator relations of the Chevalley generators
$e_i, h_i, f_i$ (where subscript $i$ is for the $\alpha_i$) are:

\be
\label{comm}
[h_i,h_j]=0, \quad [e_i,f_j]=\delta_{ij} h_j, \quad
[h_i,e_j]=A_{ij}e_j, \quad [h_i,f_i]=-A_{ij}f_j
\ee
We will use the highest weight $\vec j=\{j_1,\dots, j_r \}$
representation of $\bf g$, the
highest weight vector is denoted by $\left|\vec j,\vec 0 \right> $:

\be
\label{Ver}
\quad \left| \vec j, \vec m\right>=f_{1}^{m_1} \dots f_{r}^{m_r}
\left| \vec j,\vec 0 \right>,\quad
h_i\left| \vec j,\vec 0 \right> =2j_i\left| \vec j, \vec 0\right> ,
\quad e_i\left| \vec j,\vec 0\right> =0.
\ee
Relations (\ref{Ver}) define a
Verma module ${\mathcal V}(\vec j)$ over $\bf g$.

For the algebra $su(2)$ generators $e, f, h$ we have explicitly:

\be
\label{su2}
[h,e]=2e, \quad [h,f]=-2f, \quad [e,f]=h.
\ee
The highest weight representation with spin
$j \in {\bf N}/2$ is:

\be
\left| j, m\right> =f^m\left| j, 0 \right>, \quad
h\left| j,0 \right> =2j\left| j, 0\right>, \quad
e\left| j,0\right> =0, \quad
f^{2j+1}\left| j, 0 \right> =0.
\ee
Another useful algebra realization is given in terms of the
differential
operators acting on the polynomial ring $\Cmath [x^\a] $. The $su(2)$
realization is:

\beq
\label{difo}
\left\{
\bal
f=\d_x \\
h=2x\d_x -2j \\
e=-x^2\d_x +2jx
\eal
\right.
\eeq
The polynomial  highest weight representation
follows from the identification:

\be
\label{pol}
\left| j,m \right> \ \leftrightarrow \ \frac{x^{2j-m}}{(2j-m)!}.
\ee

\subsection{Free fields}

Let us recall some facts about free fields of use to bosonization
procedure. The first is a scalar massless bosonic field $\p$
with values in the circle, described by the action:

\be
\label{Sphi}
S_{\p}=\frac{1}{4\pi}
\int \2 \d \p \bar \d \p \ d^2 z.
\ee
With equations of motion in mind it is useful to introduce chiral
parts of
the $\p$ field:

\be
\label{pch}
\p(z,\z)=\p_{L}(z)+\p_{R}(\z)
\ee
Then we obtain the following OPE:

\be
\p_{L} (z) \p_{L}(w)=-\log(z-w)+O(1), \quad
\p_{R} (\z) \p_{R}( \w)=-\log(\z-\w)+O(1).
\ee
From (\ref{Sphi}) a useful formula for the vertex operators
correlator follows\footnote{Where $\alpha_i$ are just a complex
numbers.}:

\be
\left< \prod_{i=1}^{N}:\exp( i\alpha_i \p(z_i,\z_i )) :
\right>_{S_{\p}}=
\prod_{i<l} (z_i-z_l)^{ \alpha_i \alpha_l}
(\z_i-\z_l)^{ \alpha_i \alpha_l}
\delta_{\Sigma \alpha_i,0},
\ee
where the Cronecker's symbol comes from the
integration over the $\p$-field zero mode.
The holomorphic factorization in this expression is evident. The
constraint $\Sigma \alpha_i=0$ will be of great importance for us.
It is conventionally referred to as a neutrality condition.

The action (\ref{Sphi}) can be deformed to

\be
\label{SR}
S_{\mathcal R}=\frac{1}{4\pi} \int \left( \2 \d \p \bar \d \p +
i \a \p {\mathcal R} \sqrt{g} \right) d^2 z,
\ee
where  ${\mathcal R}$ and $g$ are the two-dimensional scalar curvature
and metric determinant correspondingly.
After choosing a special metric
$ds^2=|\omega(z)|^2$ on a sphere, where $\omega(z)$ is a meromorphic
1-differential,  the last term in the integrand
becomes proportional to the $\delta$-function.
It gives the nonzero contribution only in the singular
point $R$ of the $\omega(z)$. Thus from such a point of view the term
with the curvature results in the insertion of the vertex operator

\be
V_{vac}(R)=:\exp \left(i \a \p(R) \right):
\ee
to the point $R$ of the surface. This operator is conventionally
referred to as a "vacuum" charge~\cite{DF}.

Note that the field $\p$ in modified action (\ref{SR}) should
take values in the circle with the radius defined by identification
$\p \sim \p +2 \pi /\a$.

The second of the fields is a bosonic $\b \g$ systems with $\b$
of spin 1 and $\g$ spin 0.
There are chiral and anti-chiral versions of the corresponding action:

\be
\label{Sbg}
S_{\b \g}=\frac{1}{4 \pi} \int \b \bar \d \g \ d^2 z,
\quad
S_{\bb \bg}=\frac{1}{4 \pi} \int \bb  \d \bg \ d^2 z,
\ee
from which we read the following OPE

\be
\label{OPEbg}
\b(z) \g(w)= \frac{1}{z-w}+O(1), \quad \bb(\z) \bg(\w) =
\frac{1}{\z-\w}+O(1).
\ee
General N-point correlators in $\b \g$ system are calculated using
the Wick's
theorem and Green's functions of the $\bar \d$ and $\d$ operators
corresponding to the singular parts in the rhs of (\ref{OPEbg})

\be
\label{Nbg}
\left< \prod_{i=1}^{N} \b(z_i) \prod_{l=1}^{M} \g(w_l)
\right>_{S_{\b \g}} =
\delta_{NM} \ \sum \limits_{perm \{ \sigma\}}
\frac{1}{z_1-w_{\sigma_1}}\dots \frac{1}{z_N-w_{\sigma_N}},
\ee
and a similar one for the anti-chiral system. The neutrality
condition now is $\# \b =\# \g$.

\subsection{Free field realization of the Ka\v{c}-Moody algebras}

Let us concentrate on the holomorphic (chiral) objects.
For the sake of brevity up to section 4 we will use the notation
$\p$ for the chiral part $\p_L(z)$ of the $\p (z,\z)$.
Ka\v{c}-Moody algebra $\bf \hat g$ associated with the Lie algebra
$\bf g$ can be
described in terms of currents $J^a(z)$ with OPE

\be
J^a(z) J^b(w)=\frac{k}{2} \frac{q^{ab}}{(z-w)^2} +
\frac{f^{ab}_c}{z-w} J^c(w) +O(1),
\ee
where tensors $f^{ab}_c$ and $q^{ab}$ are structure
constants of algebra $\bf g$ and invariant Killing form.
For the case of ${\bf g}=su(2)$ they have components:

\be
q^{00}=\2q^{+-}=1, \quad f^{+-}_{0}=-2, \quad f^{0-}_{-}=-f^{0+}_{+}=1.
\ee
It is an easy exercise to check, that the currents

\be
\label{su2k}
J^+= \b \\
J^0=:\b \g:-\frac{iq}{\sqrt{2} } \d \p \\
J^-=-:\b \g^2:+i\sqrt{2} q \g \d \p +(q^2-2) \d \g
\ee
form the $su(2)_k$ algebra at the level $k=q^2$.
Note that these currents are obtained from the differential
operator representation (\ref{difo}) by the substitution

\be
\d \rightarrow \b, \quad x \rightarrow \g , \quad
j \rightarrow \frac{iq}{\sqrt{2} } \d \p
\ee
and a subsequent renormalization by adding an anomalous term
$(q^2-2) \d \g$ to the last line of the (\ref{su2k}).

Let the level $k$ be an integer. As a next step we construct a
finite-dimensional $su(2)_k$ highest weight representation with
spin $j$:

\beq
\label{V}
\left\{
\bal
V_{j,0}=\frac{\g^{2j}}{(2j)!}
:\exp \left( i \frac{\sqrt{2} j}{q} \p \right) :  \\ \\
V_{j,1}=\frac{\g^{2j-1}}{(2j-1)!}
:\exp \left( i\frac{\sqrt{2} j}{q} \p \right) :  \\ \\
\quad \quad \quad \dots \\ \\
V_{j,2j}=\ :\exp \left( i \frac{\sqrt{2} j}{ q} \p \right):
\eal
\right.
\eeq\\
The first operator in this series is the highest weight vector of the
representation, it has no singularity in the OPE with the "rising"
current
$J^-$. Acting on $V_{j,0}$ by the ``lowering" current $J^+$  one
obtains all
representation step by step.  The series is truncated because of the
vanishing factor at the singular term in the OPE \
$J^+(z)V_{j,2j}(w)$. More
explicitly,

\beq
\left\{
\bac
J^{+}(z) V_{j,m}(w)=\frac{1}{z-w} V_{j,m-1}(w) +O(1)\\ \\
J^{0}(z) V_{j,m}(w)=\frac{j-m}{z-w} V_{j,m}(w)+O(1)\\ \\
J^{-}(z) V_{j,m}(w)=\frac{m(2j-m)}{z-w} V_{j,m+1}(w)+O(1)
\eac
\right.
\eeq\\
The correspondence of (\ref{V}) with the polynomial representation
(\ref{pol})
is evident.

To study conformal properties of the vertex operators (\ref{V}),
we need a stress-tensor. It is provided by the Sugawara construction:

\be
T(z)=\frac{q_{ab}}{k+h^{\vee}} :J^a(z)J^b(z):
\ee
where $h^{\vee}$ is a dual Coxeter number (quadratic Casimir operator
in the
adjoint representation),  $h^{\vee}(SU(N))=N$, \ and $q_{ab}$ is dual
to the
$q^{ab}$. In the case of $su(2)$ one obtains by direct calculation:

\be
\label{Tsu2}
T=:\b \d \g: - \2 :(\d \p)^2:-\ \frac{i}{\sqrt{2} q} \d^2 \p.
\ee
Note that this stress-tensor corresponds to the action:

\be
\label{Schiral}
S_{chiral}=\frac{1}{4 \pi} \int \left( \2 \d \p \bar \d \p -\b
\bar \d \g +
i\frac{\sqrt{2}}{q}\p {\mathcal R} \sqrt{g} \right) d^2 z,
\ee
where the term with the curvature induces the vacuum charge

\be
V_{vac}(R)=:\exp \left(i \frac{\sqrt2}{q} \p(R) \right):
\ee

From the OPE with the stress-tensor (\ref{Tsu2}), it follows that
the conformal dimensions of all the operators from the representation
(\ref{V}) are equal to
\beq
\label{dif}
\Delta_j =\frac{j(j+1)}{k+2}.
\eeq

Besides the primary family (\ref{V}) there is one more operator
of great importance to us  --- the so-called screening
current of the conformal dimension (1,0).
The integrals of this current along the closed contours are the
screening
charges (Feigin-Fuchs operators~\cite{FFu}).
Conformal blocks are constructed as correlators of vertex
operators (\ref{V}) with the appropriate number of the screening
charges insertions~\cite{GMMOS,Dots}. The crucial property of the
screening charges is that they commute with the Ka\v{c}-Moody
currents and have zero conformal dimension.
Thus insertion of such operators does not affect the conformal and
algebraic properties of the correlator, but serves to ``screen" out
the extra charge to satisfy the neutrality condition.
For the $su(2)$ conformal blocks
we will use the following screening charge~\cite{GMMOS}:

\be
\label{Scr2}
\oint dt S(t)=\oint \b(z)
:\exp \left( -i\frac{\sqrt{2}}{q} \p(z) \right) : d z
\ee

Let us briefly discuss the free field realization for the arbitrary
Ka\v{c}-Moody algebra $\bf \hat g$. (For more details
see~\cite{PRY,PRY1}
and references therein.) One starts with the differential operator
realization of the associated Lie algebra $\bf g$ on the polynomial
ring $\Cmath [x^{\a}] $, given  by the following expressions for the
Chevalley generators:

\beq
\label{dch}
\left\{
\bal
e_{\a}(x,\d)=W_{\a}^{\b}(x) \d_{\b} \\ \\
h_i(x,\d, j)=W_{i}^{\b}(x) \d_{\b}+ \lambda_i\\ \\
f_{\a}(x,\d, j)=W_{-\a}^{\b}(x) \d_{\b}+ P^{l}_{\a}(x) \lambda_l
\eal
\right.
\eeq\\
Then one introduces a $r$ copies of the free $\p, \b, \g$\ fields
with OPE:

\be
\p_i(z) \p_j(w)=-\delta_{ij} \log (z-w), \quad
\b_i(z) \g_j(w)=\frac{\delta_{ij}}{z-w}.
\ee
After that the expressions for the corresponding Ka\v{c}-Moody
currents
are obtained  by the substitution:

\be
\d_i \rightarrow \b_i(z), \quad x_i \rightarrow \g_i(z) , \quad
\lambda_i \rightarrow {iq} \d \p_i(z)
\ee
and a subsequent renormalization by adding an anomalous term
$F^{anom}_i (\g(z), \d \g(z))$ to the lower part of the (\ref{dch}).
The primary fields are given in the terms of the following vertex
operators:

\be
\label{Vg}
V_{\vec j,\vec m}(z)=\prod \limits_{i=1}^{m}
\frac{\g_i^{2j_i-m_i}(z)}{(2j_i-m_i)!} \
:\exp \left( i \frac{\sqrt{2}}{q} \ \vec j \vec \p(z) \right) :.
\ee
The screening currents (of the first kind, in the termininology
of~\cite{PRY,PRY1}) are:

\be
S_i(z)=:W_{\a_i}^{\delta}(-\g(z))\ \b_{\delta}(z) \
\exp \left(-i \frac{\sqrt2}{q} \ \vec \a_i \vec \p(z) \right) :
\ee

\section{Conformal Blocks}

\subsection{Knizhnik-Zamolodchikov and additional algebraic equations}

It is well known, that the correlators in WZNW theory satisfy the
system of differential equations first found by Knizhnik and
Zamolodchikov\footnote{There is
an antiholomorphic system of equation as well.}~\cite{KZ}:

\be
\label{KZ0}
\left( \kappa \der{}{z_i}-\sum_{l\not =i}
\frac{t^a_it^a_l}{z_i-z_l} \right)
\left< \Phi_1 (z_1, \bar z_1) \dots \Phi_N (z_N,\bar z_N)
\right>_{WZNW}=0
\ee
where $t^a_i$ is a generator of algebra $\bf g$  that acts on
the primary field $\Phi_i$ belonging to the $i$th representation.
The summation over the $a$ indices in (\ref{KZ0}) is assumed.
The (complex)
parameter $\kappa$ is equal to $k+h^{\vee}$. Due to the holomorphic
factorization  property of the correlators in conformal field
theory~\cite{BPZ} we have:

\be
\label{hol}
\left< \Phi_1 (z_1,\z_1) \dots \Phi _N (z_N,\z_N) \right>_{WZNW}=
\sum_{a,b} C^{ab}\ {\mathcal F}_{a} (z_1, \dots, z_N)\
\overline{ {\mathcal F}_{b}(z_1, \dots, z_N)}.
\ee
where the gluing constants  $C^{ab}$ are related to the
structure constants of operator algebra.
In fact, (\ref{KZ0})  is a system
of differential equations on the so-called conformal blocks
${\mathcal F}_a(\vec z)$. From the mathematical
point of view, conformal block is a multivalued function
(to be more precise, section of a
holomorphic bundle over the moduli space of the principal
$\bf G$-bundles
over the punctured {\bf CP}$^1$) of the $N$
variables $
\vec z \equiv (z_1,\dots ,z_N)$ with values in the tensor
product of $N$
Verma modules ${\mathcal V}(\vec j_{1}) \otimes \dots \otimes
{\mathcal
V}(\vec j_{N})$ over $\bf g$.  In notation from the
subsection 2.1 the KZ equations looks like:

\be
\kappa \der{}{z_i} {\mathcal F}(\vec z)=
\sum_{j\not= i} \frac{\Omega_{ij}}{z_i-z_j} \ {\mathcal F}  (\vec z).
\ee
where

\be
\Omega_{ij}\equiv   \sum \limits_{l=1}^{r}
\ \2 \ (h_{l})_i \otimes (h_{l})_j +
\sum \limits_{\a \in \Delta_{+}}
\left( (e_{\a})_i \otimes (f_{\a})_j +
(f_{\a})_i \otimes (e_{\a})_j \right).
\ee
and $(a_{\a})_i $  stands for the element
$1 \otimes \dots a_{\a}\dots \otimes 1$  with $a_{\a}$ in the
$i$th place.

For the $su(u)$ case the additional algebraic equation is:

\be
\label{AE}
\left( \sum_{i=1}^{N} \frac{f_i}{z-z_i} \right)^{k-2j+1}
\left<\Phi_j(z) \Phi_{j_1}(z_1) \dots \Phi_{j_N}(z_N)
\right>_{WZNW}=0
\ee

\subsection{$su(2)$ case}

To represent all the primary fields from the multiplet, we will introduce
a generating function that contains all vertex operators from
the~(\ref{V}). We are interested in generating function of
a special form,
having the following OPE with the Ka\v{c}-Moody currents:

\be
\label{OPEJF}
J^{a}(z) \Phi(w)=\frac{t^a}{z-w} \Phi(w)+O(1).
\ee
This form of OPE is fixed in the bootstrap approach~\cite{KZ}.
It is easy to check that the generating function\footnote{The terms
which contain $f^n, n>2j$ vanish when $\tilde V_{j}$ acts on the vacuum
vector $\left| j,0 \right>$, so they are not important.}

\be
\label{mult}
\tilde V_{j}=\sum_{m=-\infty}^{2j} V_{j,m} f^{2j-m}
\ee
indeed has OPE (\ref{OPEJF}) with the $su(2)_k$ currents (\ref{su2k}):

\beq
\label{op}
\left\{
\bal
J^{+}(z) \tilde V_j(w) =\frac{f}{z-w} \tilde V_j(w)+O(1) \\  \\
H(z) \tilde V_j(w) =\frac{h}{z-w} \tilde V_j(w)+O(1) \\ \\
J^{-}(z) \tilde V_j(w) =\frac{e}{z-w} \tilde V_j(w)+O(1)
\eal
\right.
\eeq\\
The sum (\ref{mult}) can be rewritten in a more suitable form

\be
\label{dres}
\tilde V_j(z) \equiv \ :\exp\left(i \frac{\sqrt{2}}{q}
j \p(z) \right): \ \exp \left(\g(z) f\right),
\ee
to which we will refer as to a ``dressed" vertex operator.
Note that this form of the generating function is quite
similar to the exponential mapping from algebra to a group.

Now we are ready to calculate the conformal blocks.
The prescription is
quite simple --- we should take the holomorphic part of the
``dressed"  vertex operator $N$-point correlator with a proper
number $n$ of the screening charges:

\be
\label{bl}
{\mathcal F}_a(\vec z)=\oint \limits_{a} dt_1 \dots
dt_n \left< S(t_1)\dots S(t_n) \ \tilde V_{j_1}(z_1)\dots
\tilde V_{j_N}(z_N)
\right>_{S_{chiral}} \vec v.
\ee
where the average with the chiral action (\ref{Schiral}) is assumed.
The label $a$ marks different integration contours.
The number $n$ in this formula is dictated by the neutrality condition:

\be
\label{ch}
n=\sum_{i=1}^{N} j_i+1,
\ee
where the unity is due to the vacuum charge contribution.
Functional integral over the $\p$ field in (\ref{bl}) gives the factor

\be
\label{pr}
\prod_{p<q}(t_p-t_q)^{\frac{2}{\kappa}}
\prod_{i<l} (z_i-z_l)^{\frac{j_ij_l}{2\kappa}}
\prod_{p=1}^{n} \prod_{l=1}^{N} (t_p-z_l)^{-\frac{j_l}{\kappa}},
\ee
while the integral over the $\b \g$ fields yields

\be
\label{sum}
\sum_{\Sigma m_i=n} \frac{1}{m_1! \dots m_N!}\left< \b(t_1) \dots
\b(t_n)\  \g^{m_1}(z_1) \dots \g^{m_N}(z_N) \right> f_{1}^{m_1}
\dots f_{N}^{m_N} \ =\\
=\sum_{\Sigma m_i=n} \sum_{perm \{ \sigma \}}
\frac{f_{\sigma(1)}}{t_1-z_{\sigma(1)}} \dots
\frac{f_{\sigma(m)}}{t_m-z_{\sigma(m)}} =
\prod_{p=1}^{n} \sum_{i=1}^{N} \frac{f_i}{t_p-z_i} ,
\ee
where $\sum \limits_{perm \{ \sigma \}}$ denotes the sum over
the permutation  group of the numbers
$\{ \sigma(1), \dots, \sigma(N) \}$ such that
$\#i=m_i$ among them $(i=1,\dots,N)$.  It is useful to
introduce a notation~\cite{EFK}:

\be
\label{fun}
Y(\vec z,t)= \prod_{l=1}^{N}(t-z_l)^{-\frac{j_l}{\kappa}}
\sum_{i=1}^{N} \frac{f_i}{t-z_i}.
\ee
Then, putting (\ref{pr}) and (\ref{sum}) together we obtain:

\be
\label{lm}
{\mathcal F}_a(\vec z)=\prod_{i<l}
(z_i-z_l)^{\frac{j_i j_l}{2 \kappa}}
\oint \limits_a
dt_1 \dots dt_n \prod_{p<q}(t_p-t_q)^{\frac{2}{\kappa}} \
Y(\vec z,t_1)\dots Y(\vec z,t_n) \ \vec v.
\ee
One can check using the ``brute force" method of~\cite{SV,FSV} that
(\ref{lm}) indeed satisfies the KZ equation which for the $su(2)$
case takes
the form:

\be
\kappa \der{}{z_i} {\mathcal F}_a(\vec z)=
\sum_{j\not= i} \frac{\2 h_i h_j + e_i f_j + f_i e_j}{z_i-z_j} \
{\mathcal F}_a  (\vec z)\ ,
\ee
and the algebraic equation (\ref{AE}):

\be
\left( \sum_{i=1}^{N} \frac{f_i}{z-z_i} \right)^{k-2j+1}
{\mathcal F}_a^{(j)}  (z_0, \vec z)=0,
\ee
where ${\mathcal F}_a^{(j)}  (z_0, \vec z)$ denotes the conformal block with
insertion of the vertex operator from spin $j$ representation in the
point $z_0$.

\bigskip

{\it Polynomial representation}
\bigskip \\
Note that one can use an alternative generating
function~\cite{PRY,PRY1} which corresponds to the algebra
representation on the polynomial ring:

\be
\label{polrep}
\tilde V_{j} (z)=(1+x\g(z))^j :\exp \left(i \frac{\sqrt{2}}{q}
j \p(z) \right):.
\ee
This generating function satisfies the OPE
(\ref{op}) where $f, e, h$ generators are given by (\ref{difo}).
One can reduce the problem of the product
$\prod \limits_i (1+x_i\gamma(z_i))^{j_i}$ correlator calculation to the
problem of exponents  $\exp(x_i\g(z_i))$ correlator
calculation using the formula~\cite{PRY}:

\be
(1+x\gamma)^j =\frac{j_{\ }!}{2\pi i} \oint \limits_{u=0}du\
u^{-j-1} \exp [(1+x\gamma )u]
\ee
which follows from the Cauchy theorem. After that one obtains
the following expression for the conformal block~(\ref{bl})
in polynomial representation:

\be
{\mathcal F}_a(\vec x; \vec z)=
\left( \prod_{i<l} (z_i-z_l)^{\frac{j_i j_l}{2 \kappa}}\right)
\oint \limits_a
\left( \prod_{i=1}^{n} dt_i \right)
\left( \prod_{p<q}(t_p-t_q)^{\frac{2}{\kappa}} \right)
\left( \prod_{p=1}^{n} \prod_{l=1}^{N}
(t_p-z_l)^{-\frac{j_l}{\kappa}} \right) \times \\ \\ \times
\left( \prod_{r=1}^{N} \frac{j_r !}{2\pi i}
\oint \limits_{u_r=0} du_r\ u_{r}^{-j_r-1} e^{u_r} \right)
\ \prod_{p=1}^{n} \sum_{l=1}^{N} \frac{u_i x_i}{t_p-z_i}
\ee
However, to compare resulting conformal blocks with
the solutions of the KZ equations, obtained in~\cite{SV},
one needs to use the
first construction for the generation function.

\subsection{Simple complex Lie algebras}

As we already mentioned, generalization of this construction
to an arbitrary simple complex Lie  algebra $\bf g$ is straightforward.
One needs just to pass through three stages:

i) introduce the free chiral action:

\be
\label{Sr}
S_{r}=\frac{1}{4 \pi} \int  d^2 z
\left( \2 \d \vec \p \cdot \bar \d \vec \p -
\vec \b \bar \d \vec \g +
i \frac{ {\mathcal R} \sqrt{g}}{q} \ \vec \rho \vec\p
\right),
\ee
where  \
$ \vec \p =\{ \p_1, \dots, \p_r \}, \; r=\rank \ {\bf g}, \;
\vec a \vec b \equiv \sum \limits_{i=1}^{r} a_i b_i, \;
q=\sqrt{k+h^{\vee}}$ and $\vec \rho$ stands for the
half-sum of all positive roots.

ii) introduce the generating function:

\be
\tilde V_{\vec j}(z)=
\exp \left(\vec \g(z) \vec f_L \right) \
:\exp \left( i\frac{\sqrt2}{q}\ \vec j \vec \p(z) \right) :
\ee

iii) introduce the screening currents

\be
S_i(z)=:W_{\a_i}^{\delta}(-\g(z))\ \b_{\delta}(z) \
\exp \left(-i \frac{\sqrt2}{q} \ \vec \a_i \vec \p(z) \right) :
\ee
As a result for the algebra $\bf g$ conformal blocks one obtains:

\be
{\mathcal F}_a^{\ \bf g} (\vec z)=
\oint \limits_{a}
 d \vec t \;
\left< \prod_{i=1}^r \left( S_i(t_i[1])
\dots S_i(t_i[k_i]) \right) \
\tilde V_{\vec j_1}(z_1)\dots
\tilde V_{\vec j_N}(z_N)
\right>_{S_{r}} \vec v,
\ee
for some $\{ k_1, \dots, k_r \}$ defined by a neutrality condition.
After functional integration
the expression which coincides with the
Schechtman-Varchenko solution arises
(see also~\cite{Awa,FFR}).

\section{Correlators}

\subsection{$SU(2)$ case}

Now we turn to the problem of constructing correlators
$
\left< \Phi_1 (z_1,\z_1) \dots \Phi _N (z_N,\z_N) \right>_{WZNW}.
$\\
In more detail the  spinless primary
field multiplet $\Phi_{\Delta_j}(z,\z)$ corresponding
to the conformal dimension $\Delta_j=\frac{j(j+1)}{k+2}$\ looks like

\be
\Phi_{\Delta_j}(z,\z) =\sum_{m,\bar m=0}^{2j} c_{m} c_{\bar m}\
\Phi_{\Delta_j}^{m\bar m} (z,\z)\ f_L^m\otimes f_R^{\bar m}
\ee
with some constants $c_{m,\bar m}$ defined in such a way that
OPE (\ref{op}) for $\Phi_{\Delta_j}(z,\z)$ holds.
Upper indices $m$ and $\bar m$
are for different  primary fields from the ``left" and ``right"
multiplet  receptively.
To ``glue" correlators from the conformal blocks according to
the (\ref{hol}) we need to know the gluing constants $C^{ab}$.
They are determined by the natural physical condition on
correlators to be single-valued functions of
the primary fields complex coordinates $(z,\z)$. In mathematical
language, the monodromy  with respect to a moving point
in the correlator around the others should be trivial.
The most naive way to obtain an expression with such a property
is simply to get the ``squared modules" of the conformal blocks
and  replace the integration over contours with the
integration over the whole surface. Provided that there is a
single choice of $C^{ab}$ (this is natural from the physicist's
point of view) these prescription will give a correct answer.
Thus formally we can write:

\be
\label{cor}
\left<
\Phi_{\Delta_{j_1}} (z_1,\bar z_1)\ \dots \
\Phi_{\Delta_{j_N}} (z_N,\bar z_N)\right>_{WZNW}
\vec v \otimes \vec v \sim \\ \sim
\int   \prod_{i=1}^{n} d^2 t_i \
\left| \  \left< S(t_1)\ \dots \ S(t_n) \ \tilde V_{j_1}(z_1)\ \dots \
\tilde V_{j_N }(z_N) \right> \ \right| ^2 \ \vec v \otimes \vec v.
\ee
One can restore original conformal blocks and gluing constants
using the following formula which expresses the integral over
the Riemann surface $\Sigma$ in terms of the integrals over
canonical $A,B$ circles on this surface:

\be
\int_{\Sigma} \omega \wedge \bar \omega' =
\sum_{A,B} \left( \ \oint_A  \omega \ \overline{ \oint_B  \omega'} -
\oint_B  \omega \ \overline{\oint_A \omega'} \ \right).
\ee
where $\omega$ and $\omega'$ are arbitrary holomorphic 1-differentials.
The Riemann surface $\Sigma$ in our case
is a branch covering of {\bf CP}$^1$ defined by the multivalued
form (\ref{pr}) connected with the proper conformal block.
Corresponding methods of calculation are in fact well known from
the Dotsenko-Fateev~\cite{DF,Dots0} representation for minimal
models where the same construction is used.
As an example of operator algebra structure constants calculation
see~\cite{DF,Andr0}.

To be more precise we should substitute the following ``dressed"
vertex operator

\be
\label{dres1}
\tilde V_j(z,\z) = \exp \left(\g(z) f_L \right) \
:\exp \left(i \frac{\sqrt{2}}{q} j \p_L(z) \right):\
:\exp \left(i \frac{\sqrt{2}}{q} j \p_R(\z) \right):\
\exp \left(\bg(\z) f_R \right)
\ee
for the ``squared modulus" of the $\tilde V_j(z)$ from (\ref{cor}).
Here $f_L$ and $f_R$ are the $su(2)$ step generators which form the
``left"  and ``right" representations.
We should also substitute  expression

\be
\Scr(t,\bar t)=\b(t) \bb(\bar t)
\exp \left(-i \frac{\sqrt{2}}{q} \p(t, \bar t) \right)
\ee
for the ``squared modulus" of the screening currents.
Therefore,

\be
\label{ex4.6}
\left<
\Phi_{\Delta_{j_1}} (z_1,\bar z_1)\ \dots \
\Phi_{\Delta_{j_N}} (z_N,\bar z_N)\right>_{WZNW}
\vec v \otimes \vec v = \\ =
\const \int  \prod_{i=1}^{n} d^2 t_i \
\left< \Scr(t_1,\bar t_1)\ \dots \ \Scr(t_n,\bar t_n) \
\tilde V_{j_1}(z_1,\z_1)\ \dots \ \tilde V_{j_N }(z_N,\z_N)
\right>_{S_0}
\vec v \otimes \vec v,
\ee
where averaging with the free action

\be
S_{0}=\frac{1}{4\pi} \int d^2z
\left\{ \2 \d \p \bar \d \p -\b \bar \d \g -\bb \d \bg +
i \frac{\sqrt{2}}{q} {\mathcal R} \sqrt{g} \p \right\}
\ee
is assumed. Remarkable, expression (\ref{ex4.6}) can be rewritten in
more simple and profound form using an old idea suggested by
A.~M.~Polyakov and developed by Dotsenko and Fateev in~\cite{DF}.
Namely, consider the model with the action
$S_{\p \b \g}'=S_{0}+S_{int}$,
where interacting term is equal~to

\be
\label{int}
S_{int}= \int  d^2z \ \b \bb \exp \left(-i \frac{\sqrt{2}}{q}
\p\right)
\ee
and is exactly marginal because of the zero conformal dimension
of the exponent. Explicitely,

\be
S_{\p \b \g}'=\frac{1}{4\pi} \int d^2z
\left\{
\2 \d \p \bar \d \p -\b \bar \d \g -\bb \d \bg +
i \frac{\sqrt2}{q} {\mathcal R} \sqrt{g} \p +
 \b \bb \exp \left(-i \frac{\sqrt2}{q} \p\right)
\right\}.
\ee
All correlators in this model are calculated by
expanding in $S_{int}$. As we know, the correlator

\be
\left< \prod_{i=1}^{n} \exp \left( i\a_l \p \right) \right>
\ee
will not vanish only if the neutrality condition
$\sum_l \a_l=0$  holds. Therefore, only one term
in the series survive, and this is exactly the term in
rhs of the (\ref{ex4.6}).
Thus, we identify the WZNW correlators of primary fields
$\Phi_{\Delta(j_i)}$ with the ``dressed" vertex operator
correlators in theory with the
action $S_{\p \b \g}'$:

\be
\label{c1}
\left< \Phi_{\Delta_{j_1}} (z_1,\z_1)\dots
\Phi_{\Delta_{j_N}} (z_N,\z_N) \right>_{WZNW}
\vec v \otimes \vec v =\\
= \int {\mathcal D}[\p, \b, \g, \bb, \bg] \ \exp(-S_{\p \b \g}') \
\tilde V_{j_1}(z_1,\z_1) \dots \tilde V_{j_N }(z_N,\z_N) \
\vec v \otimes \vec v.
\ee
Straightforward calculation gives:

\be
\left< \Phi_{\Delta_{j_1}} (z_1,\z_1)\dots
\Phi_{\Delta_{j_N}} (z_N,\z_N) \right>_{WZNW}
\vec v \otimes \vec v = \const \
\prod_{i<l} |z_i-z_l|^{\frac{j_i j_l}{\kappa}}
\int  \prod_{i=1}^{n} d^2 t_i \ \times \\ \\ \times
\left( \prod_{p<q} |t_p-t_q|^{\frac{4}{\kappa}} \right)
\left( \prod_{p=1}^{n} \prod_{i=1}^{N}
|t_p-z_i|^{-\frac{2j_i}{\kappa}} \right) \
\left( \prod_{p=1}^{n} \sum_{i=1}^{N}
\frac{(f_L)_i}{t_p-z_i} \right) \ \
\left( \prod_{q=1}^{n} \sum_{l=1}^{N}
\frac{(f_R)_l}{\bar t_q-\z_l}\right) \
\vec v \otimes \vec v.
\ee

\bigskip
{\it Polynomial representation}
\bigskip \\
One can do all the above steps using the polynomial representation
(\ref{polrep}) for the generating function. As a result
in the $SU(2)$ case the following formula  arises:

\be
\label{polcor}
\left< \Phi^{(j_1)} (x_1,\bar x_1; z_1,\z_1)\  \dots \
\Phi^{(j_N)} (x_N,\bar x_N; z_N,\bar z_N) \right>_{WZNW}= \\ \\
= \const
\left( \prod_{i<l} |z_i-z_l|^{\frac{j_i j_l}{\kappa}} \right) \int
\left( \prod_{i=1}^{n} d^2 t_i \right)
\left( \prod_{p<q} |t_p-t_q|^{\frac{4}{\kappa}} \right)
\left( \prod_{p=1}^{n} \prod_{l=1}^{N}
|t_p-z_l|^{-\frac{2j_l}{\kappa}} \right)
\times \\ \\ \times
\left( \prod_{r=1}^{N} \frac{j_r !}{2\pi i} \oint
\limits_{u_r=0} du_r\ u_{r}^{-j_r-1} e^{u_r} \right)\
\left( \prod_{s=1}^{N}
\frac{j_s !}{2\pi i} \oint \limits_{\bar u_s=0} d\bar u_s \ \bar
u_{s}^{-j_r-s} e^{\bar u_s} \right)\ \left| \
\prod_{p=1}^{n} \sum_{l=1}^{N}
\frac{u_i x_i}{t_p-z_i} \ \right|^2.
\ee

\noindent

\subsection{Simple complex Lie groups}

Generalization of the construction from previous subsection
for other groups is obvious. Let us introduce the
action

\be
S_{\vec \p \vec \b \vec \g}=\frac{1}{4\pi} \int d^2z
\left\{ \2 \d \vec \p \bar \d \vec \p -\vec \b \bar \d \vec \g -
\vec{\bb} \d \vec{\bg} +
i \frac{ {\mathcal R} \sqrt{g}}{q} \vec \rho \vec \p \; + \right. \\ \\
+ \left. \sum \limits_{i} W_{\a_i}^{\delta}(-\g)
W_{\a_i}^{\bar \delta}(-\bg)\ \b_{\delta} \bb_{\bar \delta} \
\exp \left(-i \frac{\sqrt2}{q} \ \vec \a_i \vec \p(z) \right)
\right\}
\ee
and the generating function:

\be
\tilde V_{\vec j}(z,\z)=
\exp \left(\vec \g(z) \vec f_L \right)
\exp \left( i\frac{\sqrt2}{q}\ \vec j \vec \p(z,\z) \right)
\exp \left(\vec{\bg}(\z) \vec f_R \right)
\ee
Then for N-point correlator we obtain the following
representation:

\bigskip
\be
\left< \Phi_{\Delta (\vec j_1)} (z_1,\z_1) \ \dots \
\Phi_{\Delta (\vec j_N)} (z_N,\z_N) \right>_{WZNW}
\vec v \otimes \vec v = \\ \\ =
\int {\mathcal D}[\vec \p, \vec \b, \vec \g, \vec{\bb}, \vec{\bg}]
\ \exp(-S_{\vec \p \vec \b \vec \g}) \
\tilde V_{\vec j_1}(z_1,\z_1) \dots \tilde
V_{\vec j_N }(z_N,\z_N) \ \ \vec v \otimes \vec v.
\ee

To obtain expression for the correlator in polynomial
representation on needs just to substitute exponential
generating function by the corresponding polynomial one.

\section{New integral identities}

\subsection{A simple test}

Now we do some tests on the proposed construction.
It is well known that the $SU(2)_{k=1}$ WZNW model can be
represented in terms of one $\p$ field with values in the
self-dual circle. The $su(2)_{k=1}$ currents are:

\be
J^+=e^{i \sqrt2 \p}, \quad
J_0=i \sqrt2 \d \p, \quad
J^-=e^{-i \sqrt2 \p}
\ee
The are only two non-trivial primary fields correspond to the spin
$1/2$ representation:

\be
\label{bo}
\Phi_{\uparrow}=e^{\frac{i}{\sqrt2} \p}, \quad
\Phi_{\downarrow}=e^{-\frac{i}{\sqrt2} \p}
\ee
The 4-point correlator obviously is

\be
\label{ex0}
\left< \Phi_{\uparrow} (0) \Phi_{\downarrow} (x,\bar x)
\Phi_{\uparrow} (1) \Phi_{\downarrow} (\infty) \right>_{WZNW}  \ = \
\frac{\const}{\sqrt{x \bar x (1-x)(1-\bar x)}}
\ee
From the other hand, the general expression (\ref{c1}) in this case gives:

\be
\label{ex1}
\left< \Phi_{\uparrow} (0) \Phi_{\downarrow} (x)
\Phi_{\uparrow} (1) \Phi_{\downarrow} (\infty) \right>_{WZNW} \ \sim \
|x|^{1/3} \ |1-x|^{1/3} \int d^2 t_1 \
|t_1|^{-2/3} \ |t_1-x|^{-2/3}\
|t_1-1|^{-2/3} \ \times\\
\times \int d^2 t_2 \ |t_1-t_2|^{4/3}\ |t_2|^{-2/3}
|t_2-x|^{-2/3}\ |t_2-1|^{-2/3}\ \left| \ \frac{1}{t_1(t_2-1)}+
\frac{1}{t_2(t_1-1)} \ \right|^2.
\ee
The equality of this two different expressions for the correlator
seems to be a non-trivial fact. It can be easily checked
numerically.
Let us describe some steps which reduce the
problem to the
statement that some 2-fold integrals over the surface must vanish.
This statement was checked  with the help of the {\tt Mathematica 3.0}
program.
As a first step we change the integration variables:

\be
t_1=t_{1}'x, \quad t_2=t_2'-(t_2'-1)x,
\ee
so that the rhs of (\ref{ex1}) takes the form

\be
\label{ex2}
\frac{1}{\sqrt{x \bar x (1-x)(1-\bar x)}} \
\int d^2 t_1' \ |t_1'|^{-2/3} \
|1-t_1'|^{-2/3}\ |1-xt_1'|^{-2/3} \ \times\\ \\ \times \int d^2 t_2'
\ |1-t_2'|^{-2/3}\ \left| \ 1-x\frac{t_2'+t_1'-1}{t_2'} \
\right|^{4/3}\
\left| \ 1-x\frac{t_2'-1}{t_2'} \ \right|^{-2/3} \times \\ \\ \times
\left| \ \frac{1}{t_1'(t_2'-1)}+
\frac{x(1-x)}{(t_2'-x(t_2'-1))(xt_1'-1)} \
\right|^2.
\ee
If the rhs of~(\ref{ex0}) and~(\ref{ex1}) are equal, integral
in~(\ref{ex2}) should not depend on $x$. In order to check it,
we expand this integral to the series in $x$.
As a result of the numerical calculation it turns out that at
least 8 first coefficients $c_n$ $(n \ge 1)$ in front of the $x^n$
terms in this series vanish.

\subsection{Applications}

New integral identities
arise when  an alternative way to solve the WZNW model exists.
In these identities lhs and rhs are the correlators calculated in two
different ways.  They have equal conformal, algebraic and analytical
properties.  Moreover, they obey the same differential and algebraic
equations. Thus at least at physical level we can conclude
that they should be equal.

The $SU(2)_{k=1}$ WZNW $2N$-point correlator, calculated with
help of bosonization~(\ref{bo}) is

\be
\left< \prod \limits_{i=1}^{N} \Phi_{\uparrow} (z_i,\z_i)
\prod \limits_{i=1}^{N} \Phi_{\downarrow} (w_i,\w_i)
\right>_{WZNW}\ =\
\const \frac{\prod \limits_{i<j} |z_i-z_j|^{1/2} \prod \limits_{i<j}
|w_i-w_j|^{1/2}} {\prod \limits_{i,j} |z_i-w_j|^{1/2}}
\ee
Comparing with the general expression (\ref{c1}) for this correlator
gives an identity:

\be
\left( \prod_{i<j} |z_i-z_j|^{1/3}\ |w_i-w_j|^{1/3} \right)
\left( \prod_{i,j} |z_i-w_j|^{1/3} \right)
\int \left( \prod_{p=1}^{N+1}d^2 t_i \right)
\left( \prod_{p<q} |t_p-t_q|^{4/3} \right)
\times \\ \\ \times
\left( \prod_{p=1}^{N+1}  \prod_{i=1}^{N} |t_p-z_i|^{-2/3}\
|t_p-w_i|^{-2/3}\right) \ \left| \
\sum \limits_{perm \{ \sigma \} }
\frac{1}{(t_{\sigma (1)}-z_1) \dots (t_{\sigma (N)}-z_N)}
\ \right|^2 = \\ \\ =\const \
\frac{\prod \limits_{i<j} |z_i-z_j|^{1/2} \prod \limits_{i<j}
|w_i-w_j|^{1/2}}{\prod \limits_{i,j} |z_i-w_j|^{1/2}}
\ee
Another integral identity arises when one compares the prediction
(\ref{polcor}) for the 4-point $SU(2)_k$ WZNW correlator
in polynomial representation:

\be
\left< \Phi^{(j_1)}(x_1,\bar x_1; z_1,\z_1) \dots
\Phi^{(j_4)}(x_4,\bar x_4; z_4,\z_4) \right>_{WZNW} \sim \\ \\ \sim \
\left( \prod_{i<l} | z_i-z_l|^{\frac{j_i j_l}{k+2}} \right) \int
\left( \prod \limits_{i=1}^{\Sigma j_l+1} d^2t_i\right) \
\left( \prod_{p<q} |t_p-t_q|^{\frac{4}{k+2}} \right)
\left( \prod_{p=1}^{\Sigma j_l+1} \prod_{l=1}^{4}
|t_p-z_l|^{-\frac{2j_l}{k+2}} \right) \times \\ \\ \times
\left( \prod_{r=1}^{4} \frac{j_r !}{2\pi i} \oint \limits_{u_r=0}
du_r\
u_{r}^{-j_r-1} e^{u_r} \right)\ \left( \prod_{s=1}^{4}
\frac{j_s !}{2\pi i}
\oint \limits_{\bar u_s=0} d\bar u_s \ \bar u_{s}^{-j_s-1}
e^{\bar u_s}
\right)\ \left|\ \prod_{p=1}^{\Sigma j_l+1} \sum_{i=1}^{4} \frac{u_i
x_i}{t_p-z_i}\ \right|^2 \ee
with the following
expression suggested by Zamolodchikov and Fateev~\cite{ZF}:

\be
\left< \Phi^{(j_1)}(x_1,\bar x_1; z_1,\z_1) \dots
\Phi^{(j_4)}(x_4,\bar x_4; z_4,\z_4) \right>_{WZNW}=
|x_{14}|^{4j_1} |x_{24}|^{2(j_1+j_2+j_4-j_3)} \
\times \\ \\ \times
|x_{34}|^{2(j_1+j_2+j_4-j_3)} \ |x_{32}|^{2(j_1+j_2+j_3-j_4)}\
|z_{14}|^{2\nu_1} \ |z_{24}|^{2\nu_2} \ |z_{34}|^{2\nu_3}\
|z_{32}|^{2\mu}\  U_{j_1 j_2 j_3 j_4} (x,\x;z,\z),
\ee
where

\be
\nu_1=-2 \Delta^{(j_1)}, \quad \nu_2=\Delta^{(j_2)}
-\Delta^{(j_1)}-\Delta^{(j_3)}- \Delta^{(j_4)},\\ \nu_3=\Delta^{(j_3)}
-\Delta^{(j_2)}-\Delta^{(j_1)}-\Delta^{(j_4)},\\ \mu=
\Delta^{(j_1)} +\Delta^{(j_4)}-\Delta^{(j_3)}-\Delta^{(j_2)},
\ee
and $U_{j_1 j_2 j_3 j_4} (x,\x;z,\z)$ is a function of
the projective invariants:

\be
x=\frac{x_{12}x_{34}}{x_{14}x_{32}}, \quad
\bar x=\frac{\bar x_{12}\bar x_{34}}{\bar x_{14}\bar x_{32}}, \quad
z=\frac{z_{12}z_{34}}{z_{14}z_{32}}, \quad
\z=\frac{\z_{12}\z_{34}}{\z_{14}\z_{32}},
\ee
given by the following multiple integral

\be
U_{j_1 j_2 j_3 j_4}(x,\x;z,\z)=|z|^{4\frac{j_1 j_2}{k+2}}\
|1-z|^{4\frac{j_2 j_3}{k+2}} \ N(j_1 j_2 j_3 j_4) \ \times \\
\times
\int \prod \limits_{i=1}^{2j_1} d^2t_i\
|t_i-z|^{-2 \frac{\b_1}{k+2}} \ |t_i|^{-2\frac{\b_2}{k+2}} \
|t_i-1|^{-2 \frac{\b_3}{k+2}} \ |x-t_i|^2 \ |D(t)|^{\frac{4}{k+2}}
\ee
where $N(j_1 j_2 j_3 j_4)$ is certain normalization factor
(see~\cite{ZF}),  and

\be
D(t)=\prod_{i<j} (t_i-t_j),\\
\b_1=j_1+j_2+j_3+j_4+1,\\
\b_2=k+j_1+j_2-j_3-j_4+1,\\
\b_3=k+j_1+j_3-j_2-j_4+1.
\ee
To obtain new integral identities one can also use expressions for the
$SU(N)_k$ 4-point correlators of the primary fields belonging to the
fundamental representation~\cite{KZ}. As a result proper multiple
integrals (\ref{c1}) become expressed in terms of the basic
hypergeometric functions $_2F_1$.

\section{Conclusions, Speculations and Outlook}

In this paper we considered the free field approach or
bosonization
technique for the Wess-Zumino-Novikov-Witten model with arbitrary
Ka\v{c}-Moody
algebra on genus zero Riemann surface. We show how to obtain
Schechtman-Varchenko solutions of the
Knizhnik-Za\-mo\-lod\-chi\-kov equations as certain
correlators in free chiral theory thus providing a simple
description of space of the WZNW conformal blocks.
We also propose a simple
prescription for ``gluing" correlators from the conformal blocks,
quite
similar to the Dotsenko-Fateev prescription for the minimal
models~\cite{DF}.

This construction has a simple interpretation in the
functional integral language. Namely, one can attribute the
additional marginal terms in action (screening generators)
to the functional measure by means of the corresponding
$\delta$-function insertions. Thus WZNW
model becomes essentially embedded in a certain subsector of the free
fields theory. It is worth mentioning the analogy of such
description with the
powerful ``projection method" from the integrable systems
theory~\cite{OP}.

Interesting insight on bosonized action arises from the
non-critical
string theory. Let us discuss the simplest case.
Consider the bosonic Polyakov string propagating in two dimensions.
For many reasons it is useful to compactify it in a certain way.
For instance, one can consider as a target space $\bf CP^1$ with the
complex coordinates $(\g, \bg)$.  The conformal anomaly results in metric
dependence governed by the Liouville action --- this is the way how
the $\p$
field arises.  All fields become ``dressed" by the Liouville field ---
they acquire factors like $e^{\a \p}$ where $\a$ defines the anomalous
dimension.  At the moment the possible dynamics of such a non-perturbative
process is absolutely unknown. But if we are interested in the case
when the
resulting string theory has a current algebra on the world-sheet, the
form of the possible terms in effective action is strongly constrained.
For instance, the Liouville interaction term is forbidden (!) since
it results in screening charge which does not commute with the currents.
In fact, the only possibility is the action $S_{\p \b \g}'$ (\ref{S'})
corresponding to the $SU(2)$ theory or ``conjugated" action
$S_{i\p \b \g}'$ corresponding to the $SL(2,{\bf C})$ theory.
(Note that the $\b$ field can be easily integrated out.)
Of course, this is nothing but speculation. However,
it gives us an alternative way to think about our construction
and brings an interesting link to such a long-standing problem of
mathematical  physics as Liouville theory.

It is worth mentioning renewed interest to the bosonization
of the WZNW models related to the strings propagating
on $AdS_{3} \times S^{3}, AdS_{2d+1}
\times S^{2d'+1}$~\cite{BS, AdS, Andr}.
This interest is motivated by dualities between certain CFT's and string
theories on anti-de-Sitter spaces with RR fluxes~\cite{AdS/CFT}.

An intriguing direction for future research
comes from the observation~\cite{SV} that KZ equations and their
SV solutions
can be generalized to the case when $\bf g$ itself is an arbitrary
Ka\v{c}-Moody algebra associated with the symmetriable Cartan matrix.
These
generalized KZ equations should correspond to the loop group WZNW
model
while generalized  conformal blocks should correspond to the
some two-loop algebra representations. Such objects are very
important
for the unification of conformal and two-dimensional
integrable models as points on the String Theory configuration
space~\cite{GLM}.

We also believe that the tools described in the paper will help us
to understand deeper and generalize the intriguing interplay
between Langlands duality and Sklyanin's separation of
variables~\cite{FFR,Langl}.

To summarize, we suggest the following prescription,
which generalizes that one from~\cite{GMMOS,DF}:
{\it N-point
correlators of the spinless primary fields in the genus zero
WZNW model coincide with the N-point correlators of the
``dressed"  vertex operators in theory of free
($\p, \b, \g$)-fields perturbed
by the exactly marginal terms corresponding to the
``squared modules" of simple screening currents.}
As one of the possible applications of this prescription, we have
obtained a  set of new integral identities
between (naively) different hypergeometric functions.

Further development of these methods and generalization of the proposed
construction to the higher genus case will appear elsewhere.

\bigskip

\bigskip

\centerline{\large \bf Acknowledgements}

\bigskip

I am grateful to A. Marshakov and A. Mironov for
critical comments and deeply indebted to A. Morozov and A. Losev
for illuminating discussions and friendly guidance throughout the
work. I wish to thank Dima Lyubshin and Ira Vashkevich for
technical support. The work was partly supported by the Russian
President's grant 96-15-9639 and RFBR grant 98-02-16575.


\newpage

\begin{thebibliography}{99}

\bibitem{Nov} S.~Novikov, Usp.\ Mat.\ Nauk {\bf 37} (1982) 3.

\bibitem{Wit} E.~Witten,
``Non-Abelian Bosonization in Two Dimensions,"\\
Comm. Math. Phys.  {\bf 92} (1984) 455.

\bibitem{PW} A.~Polyakov and P.~Wiegmann
``Theory of Non-Abelian Goldstone Bosons,"
Phys. Lett. {\bf 131B} (1983) 121 ;
``Goldstone Fields in Two-Dimensions with Multivalued Actions,"
Phys. Lett. {\bf 141B} (1984) 223.

\bibitem{MS} G.~Moore and N.~Seiberg,
``Taming the Conformal Zoo,"
Phys. Lett. {\bf B220} (1989) 422.

\bibitem{KZ} V.~Knizhnik and A.~Zamolodchikov,
``Current Algebra and Wess-Zumino Model in Two Dimensions,"
Nucl.  Phys. {\bf B247} (1984) 83.

\bibitem{ZF} A.~Zamolodchikov and V.~Fateev,
``Operator Algebra and Correlation Functions in the Two-Dimensional
Wess-Zumino $SU(2) \times SU(2)$ Chiral Model,"
Sov. J. Nucl. Phys. {\bf 43} (1986) 657,
Yad. Fiz. {\bf 43} (1986) 1031.

\bibitem{CFlu} P.~Christe and R.~Flume,
``The Four Point Correlations Of All Primary Operators Of  The D = 2
Conformally Invariant $SU(2)$ Sigma Model With Wess-Zumino Term,''
Nucl.\ Phys.\ {\bf B282} (1987) 466.

\bibitem{BF} D.~Bernard and G.~Felder,
``Fock Representations And BRST Cohomology In $Sl(2)$ Current
Algebra,''
Commun.\ Math.\ Phys.\ {\bf 127} (1990) 145.

\bibitem{GMMOS} A.~Gerasimov, A.~Marshakov,
A.~Morozov, M.~Olshanetsky and S.~Shatashvili,
``Wess-Zumino-Witten Model as a Theory of Free Fields",
Int. J. Mod. Phys.  {\bf A5} (1990) 2495.

\bibitem{Dots} V.~Dotsenko,
``The free field representation of the $SU(2)$ conformal field
theory,"
Nucl. Phys. {\bf B338} 747 (1990);
``Solving the $SU(2)$ conformal field theory with the Wakimoto free
field representation,''
Nucl.\ Phys.\ {\bf B358} (1991) 547 .

\bibitem{Awa} H.~Awata, A.~Tsuchiya and Y.~Yamada,
``Integral Formulas for the WZNW Correlation Functions,''
Nucl.\ Phys.\ {\bf B365} (1991) 680;
H.~Awata,
``Screening Currents Ward Identity and Integral Formulas for the WZNW
Correlation Functions,''
Prog.\ Theor.\ Phys.\ Suppl.\ {\bf 110}  (1992) 303
{\tt hep-th/9202032}.

\bibitem{FGPP} P.~Furlan, A.~Ganchev, R.~Paunov and V.~Petkova,
``Reduction of the rational spin $sl(2,C)$ WZNW conformal theory,''
Phys.\ Lett.\ {\bf B267} (1991) 63;
``Solutions of the Knizhnik-Zamolodchikov equation with rational
isospins and the reduction to the minimal models,''
Nucl.\ Phys.\ {\bf B394} (1993) 665 {\tt hep-th/9201080};
A.~Ganchev and V.~Petkova,
``Reduction of the Knizhnik-Zamolodchikov equation:
A Way of producing Virasoro singular vectors,''
Phys.\ Lett.\ {\bf B293} (1992) 56 {\tt hep-th/9207032};

\bibitem{Wak} M.~Wakimoto,
``Fock Representations of the Affine Lie Algebra  $A_1^{(1)}$,"
Comm. Math Phys. {\bf 104} (1986) 605 .

\bibitem{AS}  A.~Alekseev and S.~Shatashvili,
``Path Integral Quantization of the Coadjoint Orbits of the
Virasoro Group and 2-D Gravity,"
Nucl. Phys. {\bf B323} (1989) 719 ;
``From Geometric Quantization To Conformal Field Theory,''
Commun.\ Math.\ Phys.\ {\bf 128} (1990) 197;
``Quantum Groups And WZW Models,''
Commun.\ Math.\ Phys.\ {\bf 133} (1990) 353.

\bibitem{FFr} B.~Feigin, E.~Frenkel,
``The family of representations of affine Lie algebras,"
Usp.\ Mat.\ Nauk.\ {\bf 43} (1988) 227-228,
Russ.\ Math.\ Surv.\ {\bf 43} (1989)  221--222;
``Affine Ka\v{c}-Moody algebras and semi-infinite flag manifolds,"
Commun.\ Math.\ Phys.\ {\bf 128} (1990) 161--189 ;
``Representations of affine Ka\v{c}-Moody algebras, bosonization and
resolutions,"
Lett.\ Math.\ Phys.\ {\bf 19}  (1990) 307-317;
``Representations of affine Ka\v{c}-Moody algebras and bosonization,"
pp.\ 271--316 in: Physics and Mathematics of Strings,
eds.~L.\ Brink at al, World Scientific, Singapore, 1990;
E.~Frenkel, ``Free field realizations in representation theory and
conformal field theory,"
in: Proceedings of the ICM, Z\"urich 1994, {\tt hep-th/9408109}.

\bibitem{BMcCP} P.~Bouwknegt, J.~McCarthy and K.~Pilch,
``Free Field Realizations Of WZNW Models:
BRST Complex And Its Quantum Group Structure,''
Phys.\ Lett.\ {\bf B234} (1990) 297;
``Quantum Group Structure In The Fock Space Resolutions Of $Sl(N)$
Representations,''
Commun.\ Math.\ Phys.\ {\bf 131} (1990)  125;
``Free Field Approach To Two-Dimensional Conformal Field Theories,''
Prog.\ Theor.\ Phys.\ Suppl.\ {\bf 102} (1990) 67;
``Some aspects of free field resolutions in 2-D CFT with application
to the quantum Drinfeld-Sokolov reduction,'' {\tt hep-th/9110007}

\bibitem{PRY} J.~Petersen, J.~Rasmussen and M.~Yu,
``Free field realization of $SL(2)$ correlators for admissible
representations, and hamiltonian reduction for correlators,'' Nucl.\
Phys.\ Proc.\ Suppl.\ {\bf 49} (1996) 27 {\tt hep-th/9512175};
``Conformal blocks for admissible representations in $SL(2)$ current
algebra,'' Nucl.\ Phys.\ {\bf B457} (1995) 309 {\tt hep-th/9504127,
hep-th/9510059};
J.~Rasmussen, ``Applications of free fields in 2D current algebra,''
PhD Thesis, {\tt hep-th/9610167}.

\bibitem{PRY1} J.~Petersen, J.~Rasmussen and M.~Yu,
``Free field realizations of 2D current algebras, screening currents
and  primary fields,''
Nucl.\ Phys.\ {\bf B502} (1997) 649 {\tt hep-th/9704052}.

\bibitem{DF} V.~Dotsenko and V.~Fateev,
``Conformal Algebra and Multipoint Correlation Functions in
2D Statistical Models,''
Nucl.\ Phys.\ {\bf B240} (1984) 312;
``Four Point Correlation Functions and the Operator Algebra in the
Two-Dimensional Conformal Invariant Theories with the Central Charge
$c<1$,'' Nucl.\ Phys.\ {\bf B251} (1985) 691.

\bibitem{Gaw} K. Gawedzki,
``Quadrature of Conformal Field Theories,"
Nucl. Phys. {\bf B328} (1989) 733;
``Constructive Conformal Field Theory,''
{\it In *Karpacz 1989, Proceedings, Functional integration,
geometry and strings* 277-302.},
``Geometry of Wess-Zumino-Witten models of conformal field theory,''
Nucl.\ Phys.\ Proc.\ Suppl.\ {\bf 18B} (1991) 78;
F.~Falceto, K.~Gawedzki and A.~Kupiainen,
``Scalar product of current blocks in WZW theory,''
Phys.\ Lett.\ {\bf B260} (1991) 101.

\bibitem{BPZ} A.~Belavin, A.~Polyakov and A.~ Zamolodchikov,
``Infinite Conformal Symmetry in Two-Dimensional Quantum Field
Theory,"
Nucl. Phys. {\bf B241} (1984) 333.

\bibitem{SV} V.~Schechtman and A.~Varchenko,
``Hypergeometric Solutions of Knizhnik-Zamolodchikov Equations,"
Lett.  Math. Phys.  {\bf 20} (1990) 279;
``Arrangements of Hyperplanes and Lie Algebra Homology,"
Invent.  Math., {\bf 106} (1991) 139.

\bibitem{FFu} B.~Feigin and D.~Fuks,
``Verma Modules Over the Virasoro Algebra,"
Funct. Anal. Appl. {\bf 17} 241 (1983).

\bibitem{FSV}  B.~Feigin, V.~Schechtman and A.~Varchenko,
``On Algebraic Equations Satasfied by Hyperheometric Correlators in
WZW Models. I.,"
Comm. Math. Phys. {\bf 163} (1994)  173;
``On algebraic equations satisfied by hypergeometric correlations
in WZW models. 2,''
Commun.\ Math.\ Phys.\ {\bf 170} (1995) 219 {\tt hep-th/9407010}.

\bibitem{EFK} P.~Etingof, I.~Frenkel and A.~Kirillov Jr.,
{\it Lectures on representation theory and Knizhnik-Zamolodchikov
equations,} AMS, 1998.

\bibitem{FFR} B.~Feigin, E.~Frenkel and N.~Reshetikhin,
``Gaudin model, Bethe ansatz and correlation functions at the
critical level,''
Commun.\ Math.\ Phys.\ {\bf 166} (1994) 27 {\tt hep-th/9402022};

\bibitem{Dots0} V.~Dotsenko,
``Lectures on Conformal Field Theory,"
Adv. Stud. in Pure Math. {\bf 16} (1988)~123.

\bibitem{Andr0}  O.~Andreev,
``Operator Algebra of the SL(2) Conformal Field Ftheories,"
Phys.\ Lett.\ {\bf B363} (1995) 166, {\tt hep-th/9504082.}

\bibitem{OP} M.~Olshanetsky and A.~Perelomov,
Inv.\ Math.\ {\bf 31} (1976) 93.



\bibitem{BS} J.~de Boer and S.~Shatashvili,
``Two-Dimensional Conformal Field Theories on $AdS_{2d+1}$
Backgrounds,"
{\tt hep-th/9905032}.

\bibitem{AdS} N.~Berkovits, C.~Vafa and E.~Witten,
{\tt hep-th/9902098},
D.~Kutasov and  N.~Seiberg, {\tt hep-th/9903219},
A.~Giveon, D.~Kutasov and N.~Seiberg,  {\tt  hep-th/9806194},

\bibitem{Andr} O.~Andreev,
``Unitary Representations of Some Infinite Dimensional Lie
Algebras Motivated by String Theory on AdS$_3$," {\tt hep-th/9905002};
``On Affine Lie Superalgebras, AdS$_3$/CFT Correspondence And
World-Sheets For World-Sheets,"
Nucl.Phys. {\bf B552 }(1999) 169-193 {\tt  hep-th/9901118};
``Probing AdS$_3$/CFT Correspondence via World-Sheet Methods
and 2d Gravity Like Scaling Arguments, {\tt hep-th/9909222}.

\bibitem{AdS/CFT} J.~Maldacena,
``The Large $N$ Limit of Superconformal Field Theories and
Supergravity,''
Adv. Theor. Math. Phys. {\bf 2} (1998) 231
{\tt hep-th/9711200};
S.~Gubser, I.~Klebanov and A.~Polyakov,
``Gauge Theory Correlators from Non-Critical String Theory,''
{\tt hep-th/9802109};
E.~Witten,
``Anti De Sitter Space and Holography,'' {\tt hep-th/9802150.}


\bibitem{GLM} A.~Gerasimov, D.~Lebedev and A.~Morozov,
``On Possible Implications of Integrable Systems for String Theory,''
Int.\ J.\ Mod.\ Phys.\ {\bf A6} (1991) 977.

\bibitem{Langl} E.~Frenkel,
``Affine Algebras, Langlands Duality and Bethe Ansatz,''
{\tt q-alg/9506003};
B.~Enriquez, B.~Feigin and V.~Rubtsov,
``Separation of variables for Gaudin-Calogero systems,''
{\tt q-alg/9605030};
A.~Gorsky, N.~Nekrasov and V.~Rubtsov,
``Hilbert schemes, separated variables, and D-branes,''
{\tt hep-th/9901089}.


\end{thebibliography}
\end{document}